\documentclass[10pt,letterpaper]{article}
\usepackage{opex3}
\usepackage{graphicx}

\begin{document}

\title{High power, continuous-wave, single-frequency fiber amplifier at 1091\,nm and frequency doubling to 545.5\,nm}

\author{M. Stappel, R. Steinborn, D. Kolbe, and J. Walz}

\address{Institut f$\ddot{u}$r Physik, Johannes Gutenberg-Universit$\ddot{a}$t Mainz and Helmholtz-Institut Mainz, D-55099 Mainz, Germany}

\email{stappel@uni-mainz.de}

\begin{abstract} We present a high power single-frequency ytterbium fiber amplifier system with an output power of 30\,W at 1091\,nm. The amplifier system consists of two stages, a preamplifier stage in which amplified spontaneous emission is efficiently suppressed ($>$40\,dB) and a high power amplifier with an efficiency of 52\,\%. Two different approaches of frequency doubling are compared. We achieve 8.6\,W at 545.5\,nm by single-pass frequency doubling in a MgO-doped periodically poled stoichiometric LiTaO$_{3}$ and up to 19.3\,W at 545.5\,nm by frequency doubling with a lithium-triborate (LBO) crystal in an external enhancement cavity.\end{abstract}



\section{Introduction}

Ytterbium fiber amplifiers pumped at 975\,nm are excellent sources to provide amplification in the range 1000-1150\,nm \cite{paschotta1997}. The highest output powers above the 100\,W level have been realized in the range 1040-1085\,nm \cite{gray2007,jeong2005,hildebrandt2007,mermelstein2007}. The main challenge for high power operation is to overcome amplified spontaneous emission (ASE), which competes with the signal wavelength and, if not sufficiently suppressed, limits the available output power. Especially for long signal wavelengths with low gain, this problem becomes increasingly serious. Efficient suppression of ASE requires careful selection of the amplifying fiber parameters (like core diameter, length and dopant concentration) and multiple amplifier stages with ASE filtering \cite{mermelstein2007,grot2003}. Furthermore, for amplification of narrow linewidth signals (particularly single-frequency operation), high intrafiber signal intensities can lead to stimulated Brillouin scattering (SBS), which also limits the available output power \cite{liem2003}. This effect is most important for the design of the final amplifier stage, which provides high output power. The selection of a large fiber core diameter reduces the intrafiber intensity and SBS can be avoided.

High power, narrow linewidth infrared laser sources enable efficient second-harmonic generation (SHG) in the green sprectral region. Lithium triborate (LBO) is the ideal nonlinear crystal to produce high power levels of green light due to its tolerances to optical powers. Placing a LBO inside a cavity a power of 134\,W at 532\,nm was achieved, only limited by available fundamental power \cite{meier2010}. In the past years periodically poled devices, especially MgO-doped periodically poled stoichiometric LiTaO$_{3}$ (MgO:PPSLT), turned out to be an attractive alternative for SHG of high power infrared laser sources near 1\,$\mu$m of wavelength \cite{tovstonog2007,sinha2008}. Their nonlinear coefficient is much higher which makes single-pass SHG feasible. This simplifies the experimental setup considerably because no external cavity and locking electronics are required. Up to 20\,W of radiation in the green have been demonstrated \cite{sinha2008}, although at these high power levels thermal issues become increasingly severe \cite{sinha2008,tovstonog2008}. 

In our group, high power laser radiation at 545.5\,nm is needed for a continuous-wave Lyman-$\alpha$ source based on four-wave sum frequency mixing in mercury vapour \cite{scheid2009,kolbe2011} for future laser cooling of trapped antihydrogen \cite{zimmermann1993} and Rydberg excitation of trapped Ca$^{+}$ ions \cite{FSK2011}. So far, a combination of a commerical fiber amplifier system and commercial frequency doubling unit was used, producing a harmonic output power of up to 4.1\,W at 545.5\,nm \cite{markert2007}.

In this paper we describe a high power, single-frequency, continuous-wave ytterbium fiber amplifier system at 1091\,nm and frequency doubling of the infrared light to 545.5\,nm. In section 2 we give a detailed description of the two fiber amplifier stages. The first amplifier stage provides up to 3\,W output power with efficient ASE suppression ($>$40\,dB) and the second stage amplifies the output power to 30\,W. In section 3 two different approaches of frequency doubling of the infrared light are compared. Single-pass SHG in a MgO:PPSLT crystal and SHG with LBO in an external cavity are investigated. A maximum harmonic output power of 8.6\,W and 19.3\,W at 545.5\,nm is obtained, respectively.

\section{Fiber amplifier system}
\begin{figure}[htb]
\centering\includegraphics[width=11cm]{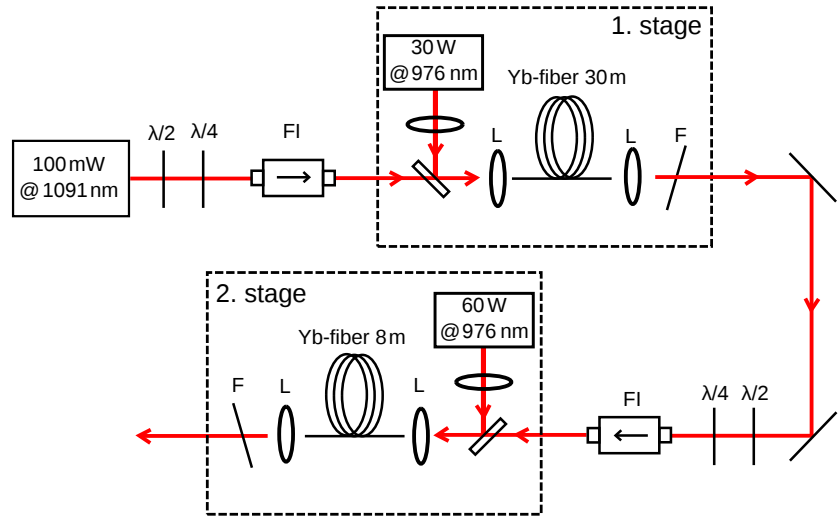}
\caption{Setup of fiber amplifier system. Light of a fiber oscillator at 1091\,nm is amplified in a first amplifier stage to medium output power and in a second amplifier to high output power. The two stages use a (free-space) forward pumping scheme. $\lambda$/2: half-wave plate, $\lambda$/4: quarter-wave plate, FI: Faraday isolator, L: lens, F: ASE-filter.}
\label{fig:setup-fibamp}
\end{figure}

Fig. \ref{fig:setup-fibamp} shows the setup of the fiber amplifier system. A single-frequency fiber laser (Koheras Adjustic Model RTAdY10PztS) with an output of 100\,mW at 1091\,nm serves as master oscillator and is amplified in two stages. Faraday isolators in front of the two amplifier stages protect the devices from backreflections (Polytec 716 TGG , transmission 80\,\%, isolation $>$60\,dB). After each stage a filter (Semrock SEM-LP02-1064RS-25, cutoff at 1075\,nm) rejects the generated ampflied spontaneous emission (ASE). We use a forward pumping scheme, where both seed and pump light are (free-space) combined by a dichroic mirror (R=99.5\,\% at 976\,nm, T=92\,\% at 1091\,nm) and coupled into the fiber.
  
\begin{figure}[htb]
\centering\includegraphics[width=13cm]{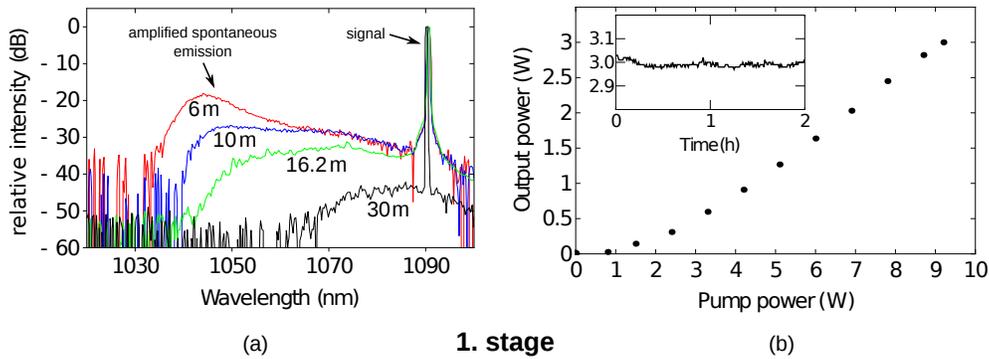}
\caption{(a) Optical output spectra for fiber lengths of 6\,m, 10\,m, 16.2\,m at a pump power of 10.5\,W and 30\,m at a pump power of 8.7\,W. Reabsorption in longer fibers reduces the amplified spontaneous emission and shifts its maximum towards longer wavelengths. (b) Output power versus pump power of the first amplifier stage. The inset shows longtime stability at maximum output power.}
\label{fig:stage1}
\end{figure}

The first amplifier stage is designed to provide medium output power with high ASE-suppression as a seed source for the second amplifier stage.
The amplifier consists of an ytterbium-doped double-clad large-mode-area fiber (Nufern, LMA-YDF-10/400, core diameter 11.5\,$\mu$m, NA 0.075, pump cladding diameter 400\,$\mu$m, NA 0.46) which maintains single-mode operation. A fiber-coupled laser diode module (Limo Lissotschenko, LIMO30-F200-DL980-T3) with up to 30 Watts at 976 nm is used to pump the active fiber. Different fiber lengths of 6\,m, 10\,m, 16.2\,m and 30\,m are tested to optimize output power and ASE-suppression. To measure the output spectrum of the ampflified light we place an optical spectrum analyzer (OSA) behind the first amplifier stage. Fig. \ref{fig:stage1}(a) shows spectral intensity versus wavelength for all tested fiber lengths. All output spectra exhibit a narrow peak at the signal wavelength, a broad maximum at smaller wavelengths and a noise floor, which corresponds to the OSA background noise. The broad maximum results from ASE and its suppression compared to the signal peak strongly depends on the fiber length. It is reduced with longer fiber length and the maximum of the ASE distribution is shifted to longer wavelengths due to reabsorption. We select the 30\,m fiber, since it provides the best ASE-supression of about 40\,dB and the highest output power. Fig. \ref{fig:stage1}(b) shows output power of the first stage as a function of pump power. We achieve a maximum output power of 3\,W corresponding to a slope efficiency of 42\,\%. This power is stable over hours as shown in the inset of Fig. \ref{fig:stage1}(b). At higher pump powers, ASE becomes to strong, which leads to self-pulsing at these wavelengths and damage of the fiber front facet occurs. To test beam quality we use a beam profiling camera (DataRay, WinCamD). A M$^{2}$ value of 1.13 was measured which is close to single-mode. This is expected for the selected fiber and ensures effective seeding of the second stage.

\begin{figure}[htb]
\centering\includegraphics[width=13cm]{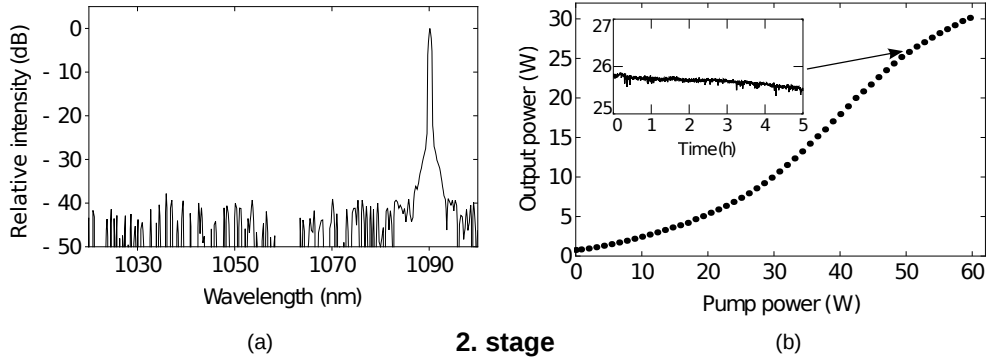}

\caption{(a) Optical output spectrum of the second amplifier stage. Amplified spontaneous emission is completely supressed. (b) Output power versus pump power of the second amplifier stage. The deviation from the linear dependency is caused by a shift in pump wavelength with pump power. The inset shows stable operation at about 26\,W over serveral hours.}
\label{fig:stage2}
\end{figure}

For the second amplifier stage we choose a 8\,m long fiber with a large core diameter of 30\,$\mu$m (Nufern, LMA-YDF-30/400, core NA 0.06) to avoid too high intensities in the fiber, which may limit the available output power as a result of stimulated Brillouin scattering \cite{liem2003}. The second stage is seeded with the radiation from the first amplifier stage, which is operated at about 2\,W. A fiber-coupled laser diode module with up to 60\,W at 976\,nm (Lumics, LU0975C060-51522A00) is used to pump the active fiber. Again, we analyze the output spectrum with an OSA. The output spectrum is shown in Fig. \ref{fig:stage2}(a) and has a peak at the signal wavelength. At all other wavelengths there is only background noise. The absence of the broad ASE maximum shows that the seed power is sufficient to suppress ASE completely. Fig. \ref{fig:stage2}(b) shows the amplified signal power plotted against the pump power. The amplified signal power deviates from the expected linear dependency. This is caused by the power dependency of the wavelength of the pump laser, which starts at 968\,nm and increases linearly to 976\,nm at maximum. At small pump powers the wavelength is far away from the absorption peak of Ytterbium at 975\,nm. Therefore only a small fraction of the coupled pump power is absorbed in the core and the efficiency is low. As the pump power increases the wavelength of the pump light comes closer to the absorption peak, more pump power is absorbed and the efficiency rises. The maximum output power of the second amplifier stage is 30\,W with an efficiency of 52\,\%, which is only limited by available pump power. For pump powers above 15\,W it is necessary to cool the connector at the front end of the fiber with a fan to prevent it from heating-up strongly. Still a constant but small decrease in output power can be observed over hours as shown in the inset of Fig. \ref{fig:stage2}(b). We assume that this is due to a self-induced process where some residual heating leads to a misalignment of the fiber, which lowers the pump light coupling efficiency and leads to more heating. However, this effect can be minimized by realigning the fiber coupling after some time of operation. We use a beam profiling camera to test the beam quality and find that it is close to single-mode (M$^{2}$ = 1.2). Although the fiber is not explicitly polarization maintaining, we observe no change of the polarization over serveral hours of operation.

\section{Frequency doubling to 545.5\,nm}
In this section we present two different approaches to frequency doubling the infrared light at 1091\,nm. First, we use a periodically poled MgO:PPSLT crystal, which has a high nonlinear coefficient and thus assures high conversion efficiencies even in a single-pass second-harmonic generation (SHG) scheme. Secondly, we investigate frequency doubling in a LBO crystal, which has a smaller nonlinear coefficient but a higher tolerance to optical powers. Therefore one can place the LBO crystal in a optical cavity, which enhances the infrared power and makes high harmonic output powers possible.  

\subsection{Single-pass second-harmonic generation in MgO:PPSLT}
\begin{figure}[htb]
\centering\includegraphics[width=11cm]{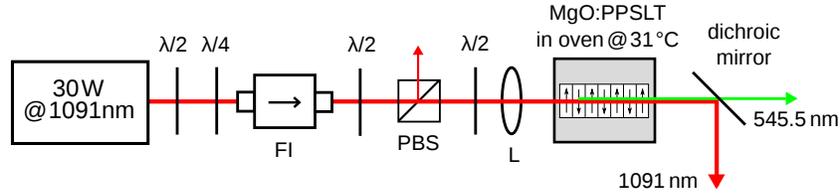}
\caption{Experimental setup of single-pass SHG in MgO:PPSLT. $\lambda$/2: half wave plate, $\lambda$/4: quarter wave plate, FI: Faraday isolator, PBS: polarizing beam splitter, L: lens}
\label{fig:setup-ppslt}
\end{figure}

The experimental setup of second-harmonic generation in the periodically poled crystal is illustrated in Fig. \ref{fig:setup-ppslt}. The fiber amplifier system described above is used as fundamental pump source. To prevent backreflections a Faraday isolator (Moltec MT-5/1091-60,  transmission 89\,\%, isolation $>$60\,dB) is placed behind the fiber amplifier. Although the Faraday isolator is specified for high power levels, we observe a considerable change in beam properties with increasing infrared power, which can be attributed to thermal lensing in the TGG (Terbium Gallium Garnet) crystal. To avoid this problem the maximum available infrared power is transmitted through the Faraday isolator and the incident power on the nonlinear crystal is adjusted by a half-wave plate and a polarizing beam splitter. An additional half-wave plate adjusts the correct polarisation for SHG.

\begin{figure}[htb]
\centering
\includegraphics[width=13cm]{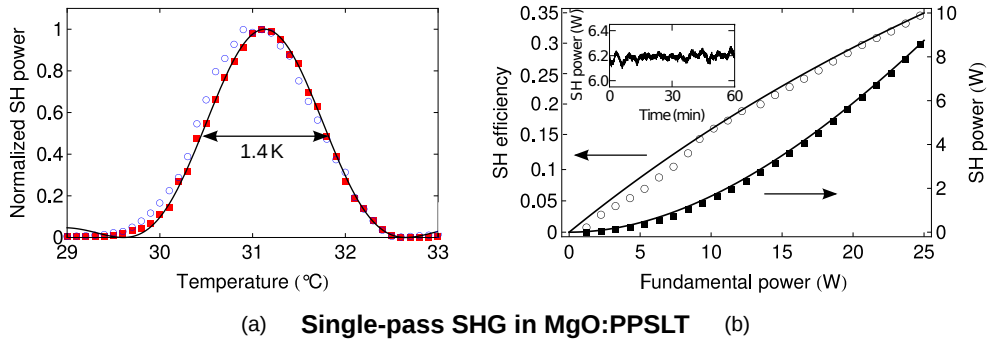}
\caption{(a) Temperature tuning curve of MgO:PPSLT for 1\,mW of green output power (red squares) and 6\,W of green output power (open blue circles). (b) Second harmonic (SH) efficiency and second harmonic power as a function of fundamental power for single-pass SHG in MgO:PPSLT.}
\label{fig:ppslt}
\end{figure}

The beam is focussed in the center of a 30\,mm long MgO:PPSLT (HC Photonics) with a beam waist radius of 34\,$\mu$m, which is slightly larger than the optimum waist radius of 29\,$\mu$m, defined by the Boyd-Kleinman parameter $l/b=2.84$ \cite{boyd1968}, where $l$ is the crystal length and $b$ is the confocal parameter. The crystal contains a single grating period of 8.61\,$\mu$m and is antireflection coated for both the fundamental (R$<$0.5\,\%) and the harmonic wavelength (R$<$1\,\%). It is mounted in a home-made oven, which is temperature stabilized with a 30x30\,mm$^{2}$ Peltier element. To ensure good heat exchange an indium foil is used between the oven and the crystal surface. The generated harmonic radiation is seperated from the fundamental radiation by a dichroic mirror (R$>$99.9\,\% at 1091\,nm, T$>$95\,\% at 545.5\,nm). 
Fig. \ref{fig:ppslt}(a) shows the temperature tuning curve for low power ($\sim$\,1\,mW) and high power ($\sim$\,6\,W) of generated harmonic radiation. The experimental data are in good agreement with the theoretical sinc$^2$-function. For the low-power measurement the phase matching temperature is 31.3\,$^\circ$C and the full width at half maximum bandwidth is 1.4\,K. For the high-power measurement the phase matching temperature is only slightly reduced to 31.0\,$^\circ$C with the same acceptance bandwidth. At high harmonic powers a deviation from the symmetric shape can occur \cite{kumar2009}. Here, at comparable powers such a behaviour can not be observed. This verifies the uniform heat distribution inside the crystal, which can be attributed to the high thermal conductivity of the MgO:PPSLT crystal and the excellent thermal contact between the crystal and the oven.

The measured second harmonic output power and the conversion efficiency as a function of incident fundamental power is shown in Fig. \ref{fig:ppslt}(b). A maximum harmonic power of 8.6\,W and an efficiency of 35\,\% is achieved. The theoretical tanh$^2$-function is matched to the experimental data (solid curve), which results in a normalized conversion efficiency of 1.8\,\%/W and an effective nonlinear coefficient of 10.9\,V/pm. The deviation from the specified value of 13.8\,V/pm can be attributed to non-ideal focussing conditions and beam quality. Even at the maximum fundamental power no saturation of conversion efficiency can be observed, indicating that thermal dephasing does not occur.
The power stability of the harmonic radiation at high power is shown in the inset of Fig. \ref{fig:ppslt}(b). A peak-to-peak fluctuation of 3\,\% is observed over 1h. To investigate the beam quality we use a beam profile camera. At all harmonic power levels M$^2$ values are $<$1.1.

\subsection{Second-harmonic generation with LBO in an external cavity}
\begin{figure}[htb]
\centering\includegraphics[width=9cm]{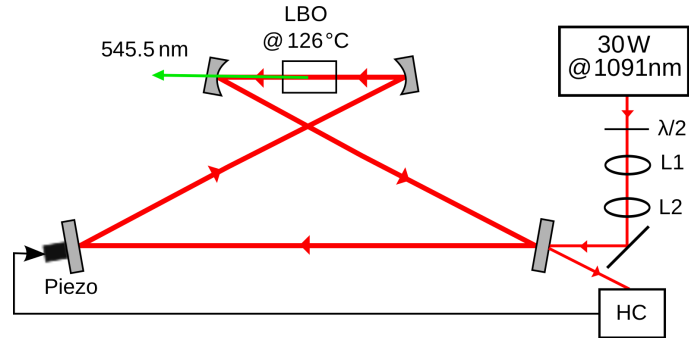}
\caption{Experimental setup of SHG in an external cavity. L1, L2: mode matching lenss, HC: H\"{a}nsch-Couillaud lock}
\label{fig:setup-cavity}
\end{figure}

In Fig. \ref{fig:setup-cavity} the experimental setup of second-harmonic generation in the external cavity is shown. Two lenses match the infrared beam to the resonator eigenmode. The external cavity is set up in the bow-tie configuration with two concave mirrors (curvature radius 50\,mm, R$>$99.9\% at 1091\,nm, T$>$95\% at 545.5\,nm), a plane mirror on a piezo actuator (R$>$99.9\% at 1091\,nm, T$>$95\% at 545.5\,nm) and a plane input coupler (R=92\,\% at 1091\,nm). Between the two concave mirrors a focus in the center of the 15\,mm long LBO crystal is generated. The LBO crystal is AR-coated for both the fundamental (R$<$0.2\,\%) and the harmonic wavelength (R$<$0.5\,\%) and temperature phase matched. The beam reflected at the input coupler is analyzed by the H\"{a}nsch-Couillaud technique and the resulting error signal is fed to the locking electronics which controls the piezo actuator. 

\begin{figure}[htb]
\centering
\includegraphics[width=13cm]{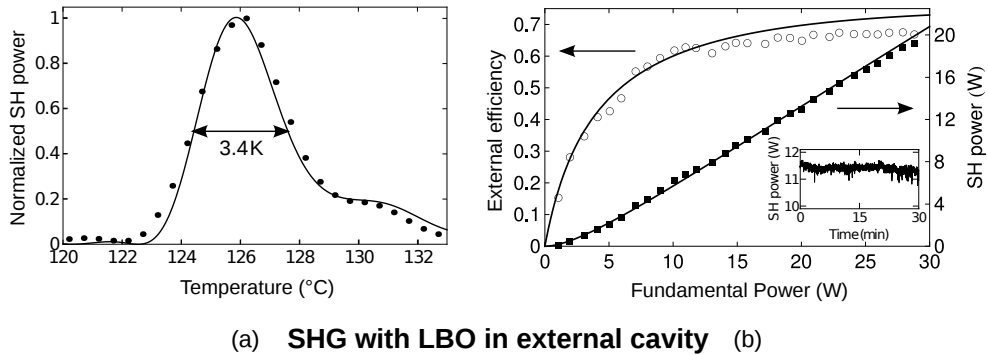}
\caption{(a) Temperature tuning curve of the LBO crystal. (b) External SH efficiency (open circles) and SH Power (black squares) as a function of fundamental power for SHG in an external cavity.}
\label{fig:lbo}
\end{figure}

Fig. \ref{fig:lbo}(a) shows the temperature tuning curve of the LBO crystal. The phase matching temperature is 126\,$^\circ$C and the temperature bandwidth is 3.4\,K. The temperature tuning curve has an asymmetric shape caused by the strong focussing of the infrared beam. We matched a theoretical curve to our data (solid curve), which is valid for focussed gaussian beams \cite{boyd1968}. For the matched curve we attain a waist radius of 15\,$\mu$m. This corresponds to a Boyd-Kleinman parameter of $l/b=7.5$, which is 2.6 times larger than the optimum value of $l/b=2.84$. As a consequence of this non-ideal Boyd-Kleinman parameter the SH efficiency is decreased by a factor of about 2. This is in agreement with the measured single-pass conversion efficiency of 0.018\,\%/W, which is half of the optimum value. However, at intra-cavity frequency doubling this parameter is less critical compared to single-pass frequency doubling, as the lower conversion efficiency is compensated to some extend by an increased enhancement of the infrared light. 
  Fig. \ref{fig:lbo}(b) shows the external conversion efficiency (second-harmonic output power divided by fundamental input power) and second harmonic output power as a function of the input fundamental power. A maximum second-harmonic power of 19.3\,W and external conversion efficiency of 67\,\% was measured. The theoretical curve is matched to the experimental data, which yields a mode-matching factor of 0.79. This is the main limiting factor of external efficiency at high incident power levels. However, the external efficiency deviates from the theoretical values, when the fundamental power is increased beyond 18\,W, and stays nearly constant for higher infrared input powers. We believe that this is due to a thermal induced change of the cavity mode leading to a decrease of the mode-matching factor. The output power was recorded over 30 minutes to test long-term stability (inset of Fig. \ref{fig:lbo}(b)). A maximum peak-peak flucatuation of 5\,\% appears and only a slight decrease of harmonic power can be observed.
To analyze the beam quality we used a beam profile camera. At all harmonic power levels M$^2$ values are $<$1.1. 

\section{Conclusion}
A high-power, continuous-wave, single-frequency ytterbium fiber amplifier system has been demonstrated, which is capable of producing 30\,W of infrared power at 1091\,nm. There is no evidence of stimulated Brillouin scattering at this power level. The signal output may thus be further enhanced by using a pump laser with more power.
 
Single-pass SHG in MgO:PPSLT yields a harmonic power of 8.6\,W. No degradation of SH efficiency at highest harmonic powers due to thermal dephasing was observed. Therefore additional fundamental power should increase the produced green power further. In comparison with frequency doubling in an external cavity, single-pass SHG is attractive, because of the experimental simplicity and the intrinsic stability.

Frequency doubling of the infrared light in an external cavity using LBO as nonlinear medium with an output of 19.3\,W at 545.5\,nm was achieved. The produced harmonic light exhibits high beam quality and longtime stability.

\section*{Acknowledgements}
This work was supported by the German Ministry for Education and Researsch (BMBF) and by the State of Rhineland-Palatinate via the Research Centre "Elementary Forces and Mathematical Foundations."

\end{document}